\documentstyle[prd,aps,epsfig]{revtex}
\begin{document}
\title{Weak form factors for heavy meson decays: an update}
\author{D. Melikhov
%\footnote{Alexander-von-Humboldt fellow; Permanent address:
%Nuclear Physics Institute, Moscow State University, 119899,
%Moscow, Russia}
and B. Stech}
\address{ITP, Universit\"at Heidelberg, Philosophenweg 16, D-69120,
Heidelberg, Germany}
\maketitle
\begin{abstract}
We calculate the form factors for weak decays of $B_{(s)}$ and $D_{(s)}$
mesons to light pseudoscalar and vector mesons. To reveal the intimate
connection between different decay modes and to be able to perform the
calculations in the full physical $q^2$-region we use a
relativistic dispersion approach based on the constituent quark picture.
This approach gives the form factors as relativistic double spectral
representations
in terms of the wave functions of the initial and final mesons. The form
factors
have the correct analytic properties and satisfy general requirements of
nonperturbative QCD
in the heavy quark limit.

The disadvantages of quark models related to ill-defined
effective quark masses and not precisely known meson wave functions
are reduced by fitting the quark model parameters
to lattice QCD results for the $B\to \rho$ transition form factors at
large momentum transfers and to the measured total $D\to (K,K^*)l\nu$
decay rates.
This allows us to predict numerous form factors for all kinematically
accessible $q^2$ values.

\vspace{.2cm}
\end{abstract}

\section{Introduction}
The knowledge of the weak transition form factors of heavy mesons is crucial
for a proper extraction of the quark mixing parameters, for the analysis of
non-leptonic decays and CP violating effects and, related to it, for a
search
of New Physics.

Theoretical approaches for calculating these form factors are
quark models \cite{wsb,isgw,isgw2,jaus,ns,orsay,faustov,gns},
QCD sum rules \cite{bbd,colangelo,lcsr,braun}, and lattice QCD
\cite{lattice,lat,ape,ukqcd}.
Although in recent years considerable progress has been made, the
theoretical
uncertainties are still uncomfortably large. An accuracy better than 15\%
has not been attained. Moreover each of the above methods has only a limited
range of applicability, namely:

QCD sum rules are suitable for describing the low $q^2$
region of the form factors. The higher $q^2$ region is hard to get and
higher
order calculations are not likely to give real progress because of the
appearance of many new parameters. The accuracy of the method
cannot be arbitrarily improved because of the necessity to isolate the
contribution of the states of interest from others.

Lattice QCD gives good results for the high $q^2$ region. But because of the
many numerical extrapolations involved this method does not provide for
a full picture of the form factors and for the relations between the various
decay channels.

Quark models do provide such relations and give the form factors in the full
$q^2$ range. However, quark models are not closely related to the QCD
Lagrangian (or at least this relationship is not well understood yet) and
therefore have input parameters which are not directly measurable and may
not be of fundamental significance.

In this situation it becomes evident that a combination of various methods
will be fruitful. It is the purpose of this paper to obtain this way
reliable predictions for many decay form factors in their full $q^2$ ranges.

To achieve this goal, one needs a general frame for the description of the
large variety of
processes. This can be only a suitable quark model, because only a quark
model
can connect different processes through the soft wave functions of the
mesons and describes the full $q^2$ range of the form factors
\footnote{One of the first steps in combining various approaches in order to
obtain
predictions for the form factors for all $q^2$ has been done in \cite{lat}
where
a simple lattice-constrained parametrization based on the constituent
quark picture of Ref. \cite{stech} and pole dominance has been proposed.}.
The predictions of this model will be much improved by incorporating the
results
from the more fundamental QCD based methods.

Historically, constituent quark models have been first to analyse the meson
transition form factors. Although a rigorous derivation of this approach as
an effective theory of QCD in the nonperturbative regime has not been
obtained,
relativistic quark models work surprisingly well for the
description of the meson spectra and form factors. Moreover, constituent
quark
models provide so far the only operative method for dealing with excited
states.

\subsubsection{The physical picture}
Constituent quark models are based on the following phenomena expected from
QCD:

i) chiral symmetry breaking in the low-energy region provides for the masses
of the
constituent quarks;

ii) the strong peaking of the soft (nonperturbative) hadronic wave functions
in terms of the quark momenta with a width of the order of
the confinement scale; and

iii) the dominance of
Fock states with the minimal number of constituents, i.e.
a $q\bar q$ composition of mesons.

An important shortcoming of previous quark model predictions was a strong
dependence of the results on the special form of the quark model
and on the parameter values.

The goal of this paper is to demonstrate that once (a) a proper relativistic
formalism is used for the description of the transition form factors and (b)
the numerical parameters of the model are chosen properly (we discuss
criteria for such
a proper choice below), the quark model yields results
in full agreement with existing experimental data and in accord with the
predictions of more fundamental theoretical approaches.
In addition, our quark model allows to predict many other form factors which
have not
yet been measured.

\subsubsection{The formalism}

For the description of the transition form factors in their full
$q^2$ range and for various initial and final mesons, a fully relativistic
treatment is necessary. We therefore choose a formulation of the quark
model which is based on the relativistic dispersion approach \cite{m1} and
thus
guarantees the correct spectral and analytic properties.

Within this model, the transition form factors are given by relativistic
double spectral representations through the wave functions of the initial
and final mesons both in the scattering and the decay regions.
These spectral representations obey rigorous constraints from QCD on the
structure of the
long-distance corrections in the heavy quark limit.
Namely, the form factors
of the dispersion quark model have the correct heavy-quark expansion at
leading and next-to-leading $1/m_Q$ orders
in accordance with QCD for transitions between heavy quarks
\cite{iwhh,luke}. For the heavy-to-light transition the dispersion quark
model
satisfies the relations between the form factors
of vector, axial-vector, and tensor currents valid at small recoil
\cite{iwhl}.
In the limit of the heavy-to-light transitions at small $q^2$ the form
factors obey the lowest order $1/m_Q$ and $1/E$ relations of the Large
Energy Effective
Theory \cite{alain} and provide the pattern of
higher-order symmetry-violating effects.

Another important advantage of the dispersion formulation of the quark model
is
the fact that one directly obtains the form factors in the physical decay
region $q^2>0$.
No numerical extrapolation from space-like values is required.

\subsubsection{Parameters of the model}

In previous applications of quark models the transition form factors
turned out to be sensitive to the numerical parameters, such as the
quark masses and the slopes of the meson wave functions.
As proposed in Ref. \cite{mns}, the way to control these parameters is to
use the results of lattice calculations at large $q^2$ as
'experimental' inputs. In \cite{mb} the $b$ and $u$ constituent quark masses
and slope parameters of the $B$, $\pi$, and $\rho$ wave functions have been
obtained through this procedure.

We now consider in addition the charm and strange mesons.
To determine the slope parameters for the charm and strange meson wave
functions and the effective mass
values $m_c$ and  $m_s$ we use here as input the measured total rates
for the decays $D\to (K, K^*)l\nu$. By fixing the parameters in this way
we overcome important uncertainties inherent in quark model calculations.
Indeed, with these few inputs we can give numerous predictions for the form
factors
for the transitions of the heavy and strange heavy mesons
$D$, $D_s$, $B$, and $B_s$ into light mesons which nicely agree at places
where data are available.

The paper is organized as follows. Section II briefly presents the main
features of the double spectral representations of the form factors.
In Section III we determine the numerical parameters of the model and give
the predictions of the form factors. Section 4 contains our conclusions.

\section{Meson transition form factors}
The long-distance contribution to meson decays is contained in the
relativistic invariant transition form factors of the vector,
axial-vector and tensor currents. The amplitudes of the $M_1\to M_2$
transition induced by the weak $q_2\to q_1$ quark transition through
the vector
$V_{\mu} = \bar q_1 \gamma_{\mu}q_2$,
axial-vector
$A_{\mu} = \bar q_1 \gamma_{\mu} \gamma^5 q_2$,
tensor
$T_{\mu\nu} = \bar q_1 \sigma_{\mu\nu} q_2$, and pseudotensor
$T^5_{\mu\nu}\;=\;{\bar q_1}\sigma_{\mu\nu}\gamma_5 q_2$ currents,
have the following covariant structure \cite{iwhl}
\begin{eqnarray}
\label{ffs}
<P(M_2,p_2)|V_\mu(0)|P(M_1,p_1)>&=&f_+(q^2)P_{\mu}+f_-(q^2)q_{\mu},
\nonumber \\
<V(M_2,p_2,\epsilon)|V_\mu(0)|P(M_1,p_1)>&=&2g(q^2)\epsilon_{\mu\nu\alpha\beta}
\epsilon^{*\nu}\,p_1^{\alpha}\,p_2^{\beta}, \nonumber \\
<V(M_2,p_2,\epsilon)|A_\mu(0)|P(M_1,p_1)>&=&
i\epsilon^{*\alpha}\,[\,f(q^2)g_{\mu\alpha}+a_+(q^2)p_{1\alpha}P_{\mu}+
a_-(q^2)p_{1\alpha}q_{\mu}\,],   \nonumber \\
<P(M_2,p_2)|T_{\mu\nu}(0)|P(M_1,p_1)>&=&-2i\,s(q^2)\,(p_{1\mu}p_{2\nu}-p_{1\
nu}p_{2\mu}), \nonumber
\\
<V(M_2,p_2,\epsilon)|T_{\mu\nu}(0)|P(M_1,p_1)>&=&i\epsilon^{*\alpha}\,
[\,g_{+}(q^2)\epsilon_{\mu\nu\alpha\beta}P^{\beta}+
g_{-}(q^2)\epsilon_{\mu\nu\alpha\beta}q^{\beta}+
g_0(q^2)p_{1\alpha}\epsilon_{\mu\nu\beta\gamma}p_1^{\beta}p_2^{\gamma}\,],
\end{eqnarray}
where $q=p_{1}-p_{2}$, $P=p_{1}+p_{2}$. The following notations are used:
$\gamma^{5}=i\gamma^{0}\gamma^{1}\gamma^{2}\gamma^{3}$,
$\sigma_{\mu \nu}={\frac{i}{2}}[\gamma_{\mu},\gamma_{\nu}]$,
$\epsilon^{0123}=-1$,
$\gamma_{5}\sigma_{\mu\nu}=-{\frac{i}{2}}\epsilon_{\mu\nu\alpha\beta}\sigma^
{\alpha\beta}$,
and
$Sp(\gamma^{5}\gamma^{\mu}\gamma^{\nu}\gamma^{\alpha}\gamma^{\beta})=
4i\epsilon^{\mu\nu\alpha\beta}$.
We study the form factors within the dispersion formulation of the quark
model \cite{m1}.
We start by considering the region $q^2<0$ where the form
factors may be represented as double spectral representations in the
invariant masses of the
initial and final $q\bar q$ pairs.
The form factors corresponding to the decay region $q^2>0$ are then derived
by performing
the analytical continuation.

\begin{figure}[hbt]
\begin{center}
\mbox{\epsfig{file=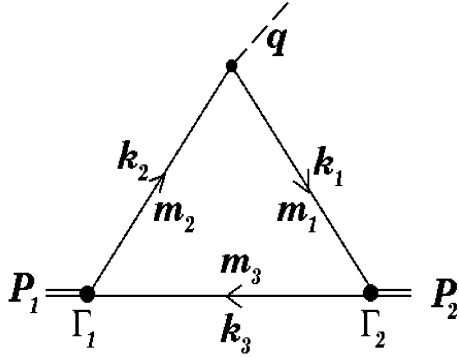,height=5.cm}  }
\end{center}
\caption{One-loop graph for a meson decay.\label{fig:trianglegraph}}
\end{figure}

The transition of the initial meson  $q(m_2)\bar q(m_3)$ with the mass $M_1$
to the final meson $q(m_1)\bar q(m_3)$ with the mass $M_2$
induced by the quark transition $q(m_2)\to q(m_1)$ through the current $\bar
q(m_1) O_\mu q(m_2)$ is described
by the diagram of Fig.\ref{fig:trianglegraph}. For constructing the double
spectral representation we must
consider a double--cut graph where all intermediate particles go on mass
shell
but the initial and final mesons have the off--shell momenta
$\tilde p_1$ and and $\tilde p_2$ such that $\tilde p_1^2=s_1$ and $\tilde
p_2^2=s_2$ with
$(\tilde p_1-\tilde p_2)^2=q^2$ kept fixed.

The quark structure of the initial and final mesons is given in terms of the
vertices $\Gamma_1$
and $\Gamma_2$, respectively. The initial pseudoscalar meson vertex has the
spinorial structure
$\Gamma_1=i\gamma_5G_1/\sqrt{N_c}$; the final meson vertex has the structure
$\Gamma_2=i\gamma_5G_2/\sqrt{N_c}$ for a pseudoscalar state and the
structure
$\Gamma_{2\mu}=[A\gamma_\mu+B(k_1-k_3)_\mu]\,G_2/\sqrt{N_c}$,
$A=-1$, $B=1/(\sqrt{s_2}+m_1+m_3)$ for an $S$--wave vector meson.

The double spectral densities $\tilde f$ of the form factors are obtained by
calculating
the relevant traces and isolating the Lorentz structures depending on
$\tilde p_1$ and $\tilde p_2$.
These invariant factors $\tilde f$ account for the two--particle
singularities in
the Feynman graph.

For $q^2<0$ the spectral representations of the form factors have the form
\cite{m1}
\begin{equation}
\label{dr}
f_i(q^2)=
\frac1{16\pi^2}\int\limits^\infty_{(m_1+m_3)^2}ds_2\varphi_2(s_2)
\int\limits^{s_1^{+}(s_2,q^2)}_{s_1^{-}(s_2,q^2)}ds_1\varphi_1(s_1)
\frac{\tilde f_i(s_1,s_2,q^2)}{\lambda^{1/2}(s_1,s_2,q^2)},
\end{equation}
where the wave function $\varphi_i(s_i)=G_i(s_i)/(s_i-M_i^2)$ and
$$
s_1^\pm(s_2,q^2)=
\frac{s_2(m_1^2+m_2^2-q^2)+q^2(m_1^2+m_3^2)-(m_1^2-m_2^2)(m_1^2-m_3^2)}{2m_1
^2}
\pm\frac{\lambda^{1/2}(s_2,m_3^2,m_1^2)\lambda^{1/2}(q^2,m_1^2,m_2^2)}{2m_1^
2}
$$
and
$
\lambda(s_1,s_2,s_3)=(s_1+s_2-s_3)^2-4s_1s_2
$
is the triangle function
\footnote{The spectral densities $\tilde f$ include proper subtraction
terms.
These subtraction terms have been determined in \cite{m1} by matching the
structure of the
heavy quark expansion in the quark model to the structure of the heavy-quark
expansion in QCD}.
The analytical continuation of the
expression (\ref{dr}) to the region $q^2>0$ gives the form factor at
$q^2\le(m_2-m_1)^2$.
Explicit expressions of the double spectral densities of all the form
factors in (\ref{ffs})
and more details can be found in \cite{m1}
\footnote{The spectral representations (\ref{dr}) take into account the
long-distance contributions connected with the structure of initial and 
final mesons. To describe additional long distance effects, Eq. (\ref{dr})
should be multiplied by the constituent quark form factor 
$f_{q_2\to q_1}(q^2)$ which contributes to the resonance structure in the
$q^2$ channel.
However, in the region of calculation $q^2<(m_2-m_1)^2$, the wave
functions provide already for a rise of the form factors with $q^2$,
which is well compatible with a properly located meson pole.
Thus, an additional quark form factor is not needed there, but we will use a
proper extrapolation formula when considering the vicinity of the poles (see
Eq (\ref{fit1}) below).}

The soft wave function of a meson $M\;[q(m_q)\bar q(m_{\bar q})]$ can
be written as follows
\begin{equation}
\label{vertex}
\varphi(s) = \frac{\pi}{\sqrt{2}} \frac{\sqrt{s^2 - (m_q^2 -
m_{\bar{q}}^2)^2}} {\sqrt{s - (m_q - m_{\bar{q}})^2}} \frac{w(k^2)}{s^{3/4}}
\end{equation}
with $k^2=\lambda(s, m_q^2, m_{\bar{q}}^2)/4s$.
Here the ground-state radial $S$-wave function $w(k^2)$
is normalized as $\int w^2(k^2)k^2 dk=1$.

As demonstrated in \cite{m1}
the form factors develop the correct structure of the long-distance
corrections
in accordance with QCD in the leading and next-to-leading $1/m_Q$
orders, if the radial wave functions $w(k^2)$ are localized in a region of
the order of the confinement scale, $k^2\lesssim\Lambda^2$.
We assume a simple gaussian parametrization of the radial wave function
\begin{equation}
\label{gauss}
w(k^2)\propto\exp(-k^2/2\beta^2),
\end{equation}
which satisfies the localization requirement for $\beta\simeq
\Lambda_{QCD}$.

The leptonic decay constant of the pseudoscalar meson $f_P$ is given in
terms
of its wave function by the spectral representation \cite{m1}
\begin{equation}
\label{fp}
f_P=\sqrt{N_c}(m_q+m_{\bar q})\int
ds\;\varphi(s)\frac{\lambda^{1/2}(s,m_q^2,m_{\bar q}^2)}
{8\pi^2 s}\frac{s-(m_q-m_{\bar q})^2}{s}.
\end{equation}

\section{Parameters of the model and numerical results}
\subsection{Parameters of the model}
We consider the slope parameter $\beta$ of the meson wave function
(\ref{gauss})
and the constituent quark masses
as fit parameters. The relevant values for the $B$, $\rho$, and $\pi$ mesons
have already been determined in \cite{mb} from a fit to the lattice results
on the form factors $T_2(q^2)$ and $A_1(q^2)$ (see Eq. (\ref{wsbff}) below)
at $q^2=19.6$ and $17.6\;$GeV$^2$ \cite{lat}.
Thereby, use has been made of the spectral representation of the leptonic
decay constant (\ref{fp}), and the double spectral representations
(\ref{dr})
of the form factors.
The values obtained for $m_b$, $m_u$, and $\beta_B$,
$\beta_\rho$, $\beta_\pi$ are displayed in Table \ref{table:parameters}.

A few comments on these numbers are in order:
\begin{itemize}
\item

The quark model double spectral representations take into account long-range
QCD effects but not the short-range perturbative corrections. However,
by fitting the wave functions and masses to reproduce the lattice points,
these
corrections are effectively taken care of:
Corrections to the quark propagators correspond to the appearance of
the effective quark masses. Corrections to the vertices at the relevant
values of the recoil variable
$\omega=(M_B^2+m_\pi^2-q^2)/2M_BM_\pi$ should be small as found in form
factors of other meson transitions \cite{amn}.
\item
The value obtained for the $b$-quark mass $m_b=4.85\;GeV$ is close to the
one-loop pole mass which in fact is the relevant mass for quark model
calculations.
\end{itemize}

We now need to fix the parameters describing the strange and charmed mesons.
The charm quark mass can be determined from the well-known $1/m_Q$ expansion
of the heavy meson mass in terms of the heavy quark
mass and the hadronic parameters
$\bar \Lambda$, $\lambda_1$ and $\lambda_2$. Using the recent estimates
of these parameters \cite{bigi} one finds
\begin{eqnarray}
m_b-m_c\simeq 3.4\; GeV.
\end{eqnarray}
This provides $m_c\simeq 1.45\;GeV$. For $m_s$ one expects $m_s\simeq
350-370\;MeV$
taking into account that $m_u=230\;MeV$.

A stringent way to constrain the parameters $m_s$, $\beta_K$, $\beta_{K^*}$,
and $\beta_D$ is provided by the measured integrated rates of the
semileptonic decays $D\to (K,K^*)l\nu$. In addition we apply the relation
(\ref{fp})
which connects $\beta_K$ with $m_s$ by using the known value of the
$K$-meson
leptonic decay constant $f_K=160\;MeV$.
The parameter values found this way are displayed in
Table \ref{table:parameters}\footnote{In \cite{m}
a different set of the parameters was used which also provided a good
description
of the available experimental data on semileptonic $B$ and $D$ decays.
However, the corresponding form factors have a rather flat $q^2$-dependence
and do not match the lattice results at large $q^2$.}.
The corresponding form
factors and decay rates are given in Tables \ref{table:fitsd2k} and
\ref{table:ratesd2k}.

The polarization of the $K^*$ in the $D\to K^*l\nu$
decays turns out to be in good agreement with the experimental result
(Table \ref{table:ratesd2k}), and
the  calculated $D$ meson decay constant $f_D=200\; MeV$ corresponds to the
expectation for the magnitude of this quantity.

The parameter $\beta_{D^*}$ cannot be found this way, but it should be
close to $\beta_D$ because of the heavy quark symmetry requirements.
We therefore set $\beta_{D^*}=\beta_D$.

Also listed in Table \ref{table:parameters} are the parameters which
describe strange heavy mesons. They are discussed in subsection D.
\begin{table}[hbt]
\caption{\label{table:parameters}
Constituent quark masses and slope parameters of the exponential wave
function
(in $GeV$).}
\centering
\begin{tabular}{|cc|c|c|c||c|c|c|c|c|c|c|c|c|c|c|}
%\hline
& $m_u$ & $m_s$ & $m_c$ & $m_b$
&  $\beta_\pi$  &$\beta_K$&  $\beta_D$ &  $\beta_B$ &
$\beta_{\eta_s}$ & $\beta_{D_s}$ & $\beta_{B_s}$ &
$\beta_\rho$ & $\beta_{K^*}$ & $\beta_{D^*}$ & $\beta_\phi$ \\
& 0.23 & 0.35  & 1.45  & 4.85  &
0.36 & 0.42  & 0.46  & 0.54  &
0.45 & 0.48  & 0.56  &
0.31 & 0.42  & 0.46  &  0.45
\end{tabular}
\end{table}
\begin{table}[hbt]
\caption{\label{table:decconst}
Leptonic decay constants of the pseudoscalar mesons in $MeV$ calculated via
\protect\ref{fp}
with the parameters of Table 1.}
\centering
\begin{tabular}{|cc|c|c|c|c|c|c|}
%\hline
&$f_\pi$  &$f_K$&  $f_D$ &  $f_B$ & $f_{\eta_s}$ & $f_{D_s}$ & $f_{B_s}$ \\
&132 & 160 & 200 & 180 &  183  &  220  &  200 \\
\end{tabular}
\end{table}
\begin{table}[hbt]
\caption{\label{table:masses}
Meson masses in $GeV$ from PDG \protect\cite{pdg}}
\centering
\begin{tabular}{|cc|c|c|c|c|c|c|c|c|c|c|c|c|}
%\hline
&$M_\pi$&$M_K$ & $M_\eta$ &$M_{\eta'}$ & $M_D$& $M_{D_s}$ & $M_B$ &
$M_{B_s}$ &
$M_\rho$&$M_{K^*}$& $M_\phi$ & $M_{D^*}$ & $M_{D^*_s}$\\
&0.14   & 0.49 & 0.547 & 0.958  & 1.87  & 1.97 & 5.27  &  5.37
&  0.77 & 0.89    &  1.02 &    2.01   &    2.11
\end{tabular}
\end{table}
The knowledge of the wave functions and the quark masses allows the
calculation of the
form factors in Eq. (\ref{ffs}). It is however more convenient to present our
results in terms
of the dimensionless form factors
$F_+$, $F_0$, $f_T$, $V$, $A_0$, $A_1$, $A_2$, $T_1$, $T_2$, $T_3$
\cite{wsb} which are the
following linear combinations of the form factors given in Eq. (\ref{ffs}):
\begin{eqnarray}
\label{wsbff}
F_+&=&f_+,\quad F_0=f_{+}+\frac{q^2}{Pq}f_{-},\quad  F_T = -(M_1+M_2)s,
\nonumber \\
V&=&(M_1+M_2)g, \quad A_1= \frac{1}{M_1+M_2}f,\quad  A_2=-(M_1+M_2)a_+,
\quad
A_0 = \frac{1}{2M_2}\left( f + q^2\cdot a_{-}+ Pq\cdot a_+\right), \nonumber
\\
T_1&=&-g_+,\quad
T_2=-g_{+}-\frac{q^2}{Pq}g_{-},\quad
T_3= g_{-}-\frac{Pq}{2} g_0.
\end{eqnarray}
The form factors (\ref{wsbff})
are defined such that they involve only contributions of resonances in the
$q^2$ channel with the same spin,
whereas some of the form factors defined by the Eq. (\ref{ffs}) contain
contributions of resonances
with different spins.
The form factors $F_+$, $F_T$, $V$, $T_1$ contain a pole at
$q^2=M_{V}^2\equiv M_{1^-}^2$
and $A_0$ contains a pole at $q^2=M_B\equiv M_{0^-}^2$ (more details are
given in the Appendix).

The remaining form factors, $F_0$, $A_1$, $A_2$, $T_2$ and $T_3$, do not
contain
contributions of the lowest lying negative parity states (for instance,
$F_0$ contains a contribution
of the $0^+$ state, and $A_1$ contains that of the $1^+$ which have
considerably
higher masses). As a result they have a rather flat $q^2$ behaviour in the
decay region, whereas
the form factors $F_+$, $F_T$, $V$, $T_1$, $A_0$ are rising  more steeply.

%\newpage

From the spectral representations (\ref{dr}) together with the parameter values
of Table I the form factors are obtained numerically. For the applications
it is convenient, however, to represent our results by simple fit formulas 
which interpolate these numerical values within a 1\% accuracy for all $q^2$
values in the region $0<q^2 < (m_2-m_1)^2$. Also, they should be appropriate
for a simple extrapolation to the resonance region.

\newpage
Let us start with the form factors $F_+$, $F_T$, $V$, $T_1$, $A_0$.
If we interpolate the results of the calculation with the simple
three-parameter fit formula
\begin{eqnarray}
\label{fit0}
f(q^2)=\frac{f(0)}{(1-q^2/M^2)(1-q^2/(\alpha M)^2)},
\end{eqnarray}
 the least-$\chi^2$
interpolation procedure leads in all cases to a value of the parameter $M$
which is within 3\% equal to the lowest resonance mass.
We consider this fact to be an important indication for the proper choice of 
the quark-model parameters and for the
reliability of our calculations. 
We therefore prefer to fix the pole mass $M$ to its physical value. 
The fit functions (\ref{fit0}) represent the results now with an accuracy of less
than 2\%. To achieve the accuracy of less than 1\% in all cases we take the form
\cite{mb}:
\begin{eqnarray}
\label{fit1}
f(q^2)=\frac{f(0)}{(1-q^2/M^2)[1-\sigma_1 q^2/M^2+\sigma_2q^4/M^4]},
\end{eqnarray}
where $M=M_P$ for the form factor $A_0$ and $M=M_V$ for the form factors
$F_+$, $F_T$, $V$, $T_1$.
In the Tables below we quote numerical values of $\sigma_2$ only if an accuracy 
of better than 1\% cannot be achieved with $\sigma_2=0$, and 
take $\sigma_2=0$ if this accuracy can already be achieved with the two 
parameters $f(0)$ and $\sigma_1$. A two-parameter fit was discussed in 
\cite{kaidalov}.

For the heavy-to-light meson transitions the masses of the lowest resonances
are not very much different from the highest $q^2$ values in the decay.
Eq. (\ref{fit1}) then allows an estimate of the residues of these poles. These
residues can be expressed in terms of products of weak and strong coupling
constants (see  Appendix). The errors for these constants induced by 
changing $\sigma_1 $ and $\sigma_2$ in our fitting procedure 
(keeping to the 1\% requirement) do not exceed 10\%. Moreover,
the residues of the form factors at the meson pole are not
independent and satisfy certain constraint (see Eq. (\ref{constraint}) in
the Appendix), which provides a consistency check of the extrapolations. 
The mismatch in (\ref{constraint}) is always below 10\%, and in most of the
cases much lower. 

For the form factors $F_0$, $A_1$, $A_2$, $T_2$ and $T_3$ the contributing
resonances ($0^+$, $1^+$, etc)
lie farther away from the physical decay region and  the effect of any
particular resonance is smeared out.
For these form factors the interpolation formula taken is\footnote{ 
One should note that the parameters $\sigma_1$ and $\sigma_2$ in 
the fit formula (\ref{fit3}) for the form factors $F_0$, $A_1$, $A_2$, $T_2$, and $T_3$
are introduced in a different way than in the fit formula (\ref{fit1}) for the form factors 
$F_+$, $F_T$, $V$, $T_1$, and $A_0$.} 
\begin{eqnarray}
\label{fit3}
f(q^2)= f(0) / [1-\sigma_1 q^2/M_V^2+\sigma_2q^4/M_V^4].
\end{eqnarray}
If setting $\sigma_2=0$ allows us to describe the calculation results
with better than
1\% accuracy for all $q^2$, a simple monopole two-parameter formula is used.

The values of $f(0)$, $\sigma_1$, and $\sigma_2$ are given for each decay
mode in the
relevant subsections.

%\newpage
%___________________________________________________________________________
%_________________
\subsection{Charmed meson decays}
\subsubsection{$D\to K,K^*$}
The $D\to K,K^*$ decays are induced by the charged current $c\to s$
quark transition. As described in the previous section, the measured total
rates of these decays are
used for a precise fit of the parameters of our model. With the parameters
of Table \ref{table:parameters}
we obtain the form factors listed in Table \ref{table:fitsd2k}.
Table \ref{table:compd2k} compares the form factors at $q^2=0$ with the
results
of other approaches and Table \ref{table:ratesd2k} presents the decay rates.

\begin{table}[hbt]
\caption{\label{table:fitsd2k}The $D \to K, K^*$ transition form
factors. $M_V=M_{D_s^*}=2.11\; GeV$, $M_P=M_{D_s}=1.97\; GeV$. 
For the form factors $F_+, F_T, V, A_0, T_1$ the fit formula Eq. (\protect\ref{fit1}) is used, 
for the other form factors - Eq. (\protect\ref{fit3})}
\centering
\begin{tabular}{|c||c|c|c||c|c|c|c|c|c|c|}
%\hline
 & \multicolumn{3}{c||}{$D \to K$} & \multicolumn{7}{c|}{$D \to K^*$}\\
\hline
          &$F_{+}$&$F_{0}$& $F_T$ & $V$   & $A_0$ & $A_1$  & $A_2$ & $T_1$
& $T_2$  & $T_3$   \\
\hline
$f(0)$     & 0.78 & 0.78  & 0.75  & 1.03  & 0.76  &  0.66  & 0.49  &  0.78
& 0.78   & 0.45    \\
$\sigma_1$ & 0.24 & 0.38  & 0.27  & 0.27  & 0.17  &  0.30  & 0.67  &  0.25
& 0.02   & 1.23    \\
$\sigma_2$ &      & 0.46  &       &       &       &  0.20  & 0.16  &
& 1.80   & 0.34     \\
\end{tabular}
\end{table}
\begin{table}[hbt]
\caption{\label{table:compd2k}
Comparison of the results of different approaches for the semileptonic
$D\to K,K^*$ form factors at $q^2=0$.}
\centering
\begin{tabular}{|c||c|c|c|c|c|}
%\hline
 Ref.                &$F_+(0)$   & $F_T(0)$  &  $V(0) $  & $A_1(0)$  &
$A_2(0)$    \\
\hline
This work
                     &  0.78     &   0.75    & 1.03      &  0.66     &  0.49
\\
WSB \cite{wsb}
                     &  0.76     &    $-$    & 1.3       &  0.88     &  1.2
\\
Jaus'96 \cite{jaus}
                     &  0.78     &    $-$    & 1.04      &  0.66     &  0.43        
\\
SR \cite{bbd}
                     &  0.60(15) &    $-$    & 1.10(25)  &  0.50(15) &0.60(15)    
\\
Lat(average) \cite{lattice}
                     &  0.73(7)  &    $-$    & 1.2(2)    &  0.70(7)  &0.6(1)    
\\
Lat \cite{ape}
                     &  0.71(3)  &  0.66(5)  &    $-$    &    $-$    &$-$     
\\
\hline 
Exp \cite{ryd}
                     &  0.76(3)  &    $-$       &  1.07(9)  &  0.58(3)  &
0.41(5)
%\\  \hline
\end{tabular}
\end{table}
\begin{table}[hbt]
\caption{\label{table:ratesd2k}
The $D\to (K,K^*)l\nu$ decay rates in $10^{10}\;s^{-1}$ obtained
within different approaches, $|V_{cs}|=0.975$.}
\centering
\begin{tabular}{|c||c|c|c|c|}
%\hline
 Ref.               &$\Gamma(D\to K)$ & $\Gamma(D\to K^*)$ &
$\Gamma(K^*)/\Gamma(K)$ & $\Gamma_L/\Gamma_T$ \\
\hline
This work           &  9.7      & 6.0       &  0.63     &  1.28       \\
Jaus'96 \cite{jaus} &  9.6      & 5.5       &  0.57     &  1.33       \\
SR \cite{bbd}       &  6.5(1.5) & 3.8(1.5)  &  0.50(15) &  0.86(6)    \\
\hline
Exp \cite{cleod2k}  &  9.3(4)   & 5.7(7)    &  0.61(7)  &  1.23(13)
\cite{pdg}
%\\ \hline
\end{tabular}
\end{table}
Extrapolating the form factors to $q^2=M^2_{D^*}$ (or $q^2=M_D^2$ for $A_0$)
gives the following estimates of the coupling constants (see the Appendix
for the relevant formulas)
\begin{eqnarray}
\frac{g_{D_s^*DK}f_V^{(D_s^*)}}{2M_{D_s^*}}=1.05\pm0.05, \qquad
\frac{g_{D_sDK^*}f_P^{(D_s)}}{2M_{K^*}}=1.7\pm 0.1, \qquad
\frac{f_T^{(D_s^*)}}{f_V^{(D_s^*)}}=0.95\pm0.05, \qquad
\frac{g_{D_s^*DK^*}}{g_{D_s^*DK}}=1.1\pm0.1. \nonumber
\end{eqnarray}
%\newpage
%___________________________________________________________________________
%_________________
\subsubsection{$D\to \pi,\rho$}
These decays are induced by the $c\to d$ charged current. Since all the
necessary parameters
have already been fixed, this mode allows for parameter-free predictions.
Table  \ref{table:fitsd2pi} presents the results of our calculations. In
Tables \ref{table:compd2pi} and
\ref{table:ratesd2pi} we compare our results with different approaches and
with experimantal data.
The form factors at $q^2=0$ are close to the
predictions of the relativistic quark model of Ref. \cite{jaus}, but the
$q^2$ dependence is different such that our model and \cite{jaus}
predict different decay rates.
Although the experimental errors are very
large and nearly all theoretical results agree with experiment, we notice
perfect agreement of our decay rates with the central values.

%%\newpage
\begin{table}[hbt]
\caption{\label{table:fitsd2pi}The calculated $D \to \pi,\rho$ transition
form
factors. $M_V=M_{D^*}=2.01\; GeV$, $M_P=M_{D}=1.87\; GeV$.
For the form factors $F_+, F_T, V, A_0, T_1$ the fit formula Eq. (\protect\ref{fit1}) is used, 
for the other form factors - Eq. (\protect\ref{fit3}).}
\centering
\begin{tabular}{|c||c|c|c||c|c|c|c|c|c|c|}
%\hline
 & \multicolumn{3}{c||}{$D \to \pi$} & \multicolumn{7}{c|}{$D \to \rho$}\\
\hline
          &$F_{+}$&$F_{0}$& $F_T$ & $V$   & $A_0$ & $A_1$  & $A_2$ & $T_1$
& $T_2$  & $T_3$   \\
\hline
$f(0)$     & 0.69 & 0.69  & 0.60  & 0.90  & 0.66  &  0.59  & 0.49  &  0.66
& 0.66   & 0.31    \\
$\sigma_1$ & 0.30 & 0.54  & 0.34  & 0.46  & 0.36  &  0.50  & 0.89  &  0.44
& 0.38   & 1.10    \\
$\sigma_2$ &      & 0.32  &       &       &       &        &       &
& 0.50   & 0.17
\end{tabular}
\end{table}

\begin{table}[hbt]
\caption{\label{table:compd2pi}
Comparison of the results of different approaches for the  semileptonic
$D\to \pi,\rho$ form factors at $q^2=0$.}
\centering
\begin{tabular}{|c||c|c|c|c|c|}
%\hline
 Ref.                    &$F_+(0)$   & $F_T(0)$  &$V(0) $   & $A_1(0)$  &
$A_2(0)$  \\
\hline
This work                &  0.69     & 0.60      &  0.90     &  0.59      &
0.49     \\
WSB  \cite{wsb}          &  0.69     & $-$       & 1.23      &  0.78     &
0.92     \\
Jaus'96\cite{jaus}       &  0.67     & $-$       & 0.93      &  0.58     &
0.42     \\
SR  \cite{bbd}           &  0.50(15) & $-$       & 1.0(2)    &  0.5(2)   &
0.4(2)   \\
Lat(ave) \cite{lattice}  &  0.65(10) & $-$       & 1.1(2)    &  0.65(7)  &
0.55(10)    \\
Lat \cite{ape}         &  0.64(5)  &  0.60(7)  &    $-$    &    $-$    &
$-$     \\
%\hline
\end{tabular}
\end{table}

\begin{table}[hbt]
\caption{\label{table:ratesd2pi}
The $D\to (\pi,\rho)l\nu$ decay rates in $10^{10}\;s^{-1}$,
$|V_{cd}|=0.22$.}
\centering
\begin{tabular}{|c||c|c|c|c|}
%\hline
 Ref.             &$\Gamma(D\to \pi)$ & $\Gamma(D\to \rho)$ &
$\Gamma(\rho)/\Gamma(\pi)$ & $\Gamma_L/\Gamma_T$ \\
\hline
This work                &  0.95     & 0.42      &  0.45     &  1.16
\\
WSB \cite{wsb}           &  0.68     & 0.67      &  1.0      &  0.91
\\
Jaus'96\cite{jaus}       &  0.8      & 0.33      &  0.41     &  1.22
\\
SR \cite{bbd}            &  0.39(8)  & 0.12(4)   &  $-$      &  1.31(11)
\\
Melikhov'97 \cite{m}     &  0.62     & 0.26      &  0.41     &  1.27
\\
\hline
Exp \cite{expd2pi}       &  0.92(45) & 0.45(22)  &  0.50(35) &  $-$
%\\  \hline
\end{tabular}
\end{table}
For the coupling constants we get the following relations
\begin{eqnarray}
\frac{g_{D^*D\pi}f_V^{(D^*)}}{2M_{D^*}}= 1.05\pm0.05, \qquad
\frac{g_{DD\rho}f_P^{(D)}}{2M_\rho}=2.1\pm0.2,  \qquad
\frac{f_T^{(D^*)}}{f_V^{(D^*)}}=0.9\pm0.1,  \qquad
\frac{g_{D^*D\rho}}{g_{D^*D\pi}}=1.3\pm0.2.  \nonumber
\end{eqnarray}
Taking $f_V^{(D^*)}\simeq 220\;MeV$, we find
$$
g_{D^*D\pi}=18\pm 3
$$
in perfect agreement with a calculation of $g_{D^*D\pi}$ based on combining
PCAC with the
dispersion approach \cite{mbg}.
%\newpage
%___________________________________________________________________________
%_________________
\subsection{Beauty meson decays}
\subsubsection{$B\to D,D^*$}
These decays arise from the heavy-quark $b\to c$ transition. Here one has
rigorous
predictions for the expansion of the form factors in terms of the
heavy-quark
mass \cite{luke}. Namely, the main part of the form factors can be expressed
through the
universal form factor - the Isgur-Wise function. However, different models
provide
different $q^2$-dependences of the Isgur-Wise function as well as
different subleading $1/m_Q$ corrections.

We recall that our spectral representations of the form factors explicitly
respect
the structure of the long-distance QCD corrections
in the leading and the subleading orders of the heavy-quark expansion. Thus,
we expect reliable predictions for the form factors. Our numerical
results are
summarized in Tables \ref{table:fitsb2d}, \ref{table:compb2d}, and
\ref{table:ratesb2d}.
\begin{table}[hbt]
\caption{\label{table:fitsb2d} The calculated $B \to D,D^*$ transition form
factors.
$M_V=M_{B_c^*}\simeq M_P=M_{B_c}=6.4\; GeV$. 
For the form factors $F_+, F_T, V, A_0, T_1$ the fit formula Eq. (\protect\ref{fit1}) is used, 
for the other form factors - Eq. (\protect\ref{fit3}).}
\centering
\begin{tabular}{|c||c|c|c||c|c|c|c|c|c|c|}
%\hline
 & \multicolumn{3}{c||}{$B \to D$} & \multicolumn{7}{c|}{$B \to D^*$}\\
\hline
          &$F_{+}$&$F_{0}$& $F_T$ & $V$   & $A_0$ &  $A_1$  & $A_2$ & $T_1$
& $T_2$  & $T_3$   \\
\hline
$f(0)$     & 0.67 & 0.67  & 0.69  & 0.76  & 0.69  &  0.66  & 0.62  &  0.68
& 0.68   & 0.33    \\
$\sigma_1$ & 0.57 & 0.78  & 0.56  & 0.57  & 0.58  &  0.78  & 1.40  &  0.57
& 0.64   & 1.46    \\
$\sigma_2$ &      &       &       &       &       &        & 0.41  &
&        & 0.50     \\
\end{tabular}
\end{table}
\begin{table}[hbt]
\caption{\label{table:compb2d}
Comparison of the results of different approaches for the semileptonic
$B\to D,D^*$ form factors at $q^2=0$.}
\centering
\begin{tabular}{|c||c|c|c|c|}
%\hline
 Ref.                 &$F_+(0)$    & $A_1(0)$  & $R_1(0)=V(0)/A_1(0)$ &
$R_2(0)=A_2(0)/A_1(0)$  \\
\hline
This work             &  0.67      &  0.66     &  1.15               &0.94  
\\
Jaus'96 \cite{jaus}   &  0.69      &  0.69     &  1.17               & 0.93
\\
Neubert \cite{neubert}&            &           &  1.3                &  0.8
\\
Close,Wambach \cite{close}&       &           &  1.15                &  0.91
\\
\hline
Exp \cite{cleoR1}     &            &           &  1.18$\pm$0.15$\pm$0.16 &
0.71$\pm$0.22$\pm$0.07
\end{tabular}
\end{table}
\begin{table}[hbt]
\caption{\label{table:ratesb2d}
The $B\to (D,D^*)l\nu$ decay rates in $|V_{cb}|^2\;ps^{-1}$. }
\centering
\begin{tabular}{|c||c|c|c|c|}
 Ref.      &$\Gamma(B\to D)$ & $\Gamma(B\to D^*)$ & $\Gamma(D^*)/\Gamma(D)$
& $\Gamma_L/\Gamma_T$ \\
\hline
This work               &  8.57    & 22.82 & 2.66     &  1.11   \\
%WSB  \cite{wsb}         &  8.1     & 21.9  & 2.71     &         \\
Jaus'96 \cite{jaus}     &  9.6     & 25.33 & 2.64     &         \\
Melikhov'97 \cite{m}    &  8.7     & 21.0  & 2.65     &  1.28   \\
\hline
Exp        &  $(1.34\pm0.15)10^{-2}\; ps^{-1}$ \cite{cleob2d}  &
$(2.98\pm0.17)10^{-2}\; ps^{-1}$ \cite{pdg}  & 2.35(1.3) &
$1.24(0.16)$  \cite{cleogr}
%\\ \hline
\end{tabular}
\end{table}
For the coupling constants we find
\begin{eqnarray}
\frac{g_{B_c^*BD}f_V^{(B_c^*)}}{2M_{B_c^*}}= 1.56\pm0.15, \qquad
\frac{g_{B_cBD^*}f_P^{(B_c)}}{2M_{D^*}}=3.3\pm0.3, \qquad
\frac{f_T^{(B_c^*)}}{f_V^{(B_c^*)}}=0.9\pm0.1, \qquad
\frac{g_{B_c^*BD^*}}{g_{B_c^*BD}}=1.05\pm0.05.
\nonumber
\end{eqnarray}
%\newpage
%___________________________________________________________________________

%_________________
\subsubsection{$B\to K,K^*$}
\begin{figure}[b]
%\vspace{1cm}
\begin{center}
\epsfig{file=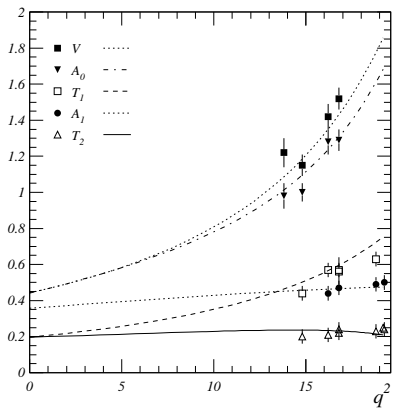,width=6cm}
%\vspace{0.5cm}
\caption{Form factors of the $B\to K^*$ transition vs the lattice results.
\label{fig:b2kstar}}
\end{center}
%\vspace{1cm}
\end{figure}

These decays are induced by the $b\to s$ Flavor Changing Neutral Current
(FCNC).
We recall that the $B\to \pi,\rho$ form factors at large $q^2$ have been
used to fix
the parameters of the model. Thus we expect that the predictions for the
$B\to K,K^*$ form factors which in fact differ from the former mode only by
SU(3) violating effects should be particularly reliable.
Table \ref{table:fitsb2k} presents the calculated form factors and Fig
\ref{fig:b2kstar}
exhibits our predictions together with the available lattice results at large
$q^2$. The
good agreement shows that the size and the sign of the SU(3) violating
effects
are correctly accounted for.
\begin{table}[hbt]
\caption{\label{table:fitsb2k}The calculated $B \to K, K^*$ transition form
factors.
$M_V=M_{B_s^*}=5.42\; GeV$, $M_P=M_{B_s}=5.37\; GeV$.
For the form factors $F_+, F_T, V, A_0, T_1$ the fit formula Eq. (\protect\ref{fit1}) is used, 
for the other form factors - Eq. (\protect\ref{fit3}).}
\centering
\begin{tabular}{|c||c|c|c||c|c|c|c|c|c|c|}
%\hline
 & \multicolumn{3}{c||}{$B \to K$} & \multicolumn{7}{c|}{$B \to K^*$}\\
\hline
          &$F_{+}$&$F_{0}$& $F_T$ & $V$   & $A_0$ &  $A_1$  & $A_2$ & $T_1$
& $T_2$  & $T_3$   \\
\hline
$f(0)$     & 0.36 & 0.36  & 0.35  & 0.44  & 0.45  &  0.36  & 0.32  &  0.39
& 0.39   & 0.27    \\
$\sigma_1$ & 0.43 & 0.70  & 0.43  & 0.45  & 0.46  &  0.64  & 1.23  &  0.45
& 0.72   & 1.31    \\
$\sigma_2$ &      & 0.27  &       &       &       &  0.36  & 0.38  &
& 0.62   & 0.41
 \end{tabular}
\end{table}
\begin{table}[hbt]
\caption{\label{table:compb2k}
Comparison of the results of different approaches for the
$B\to K,K^*$ form factors at $q^2=0$.}
\centering
\begin{tabular}{|c||c|c|c|c|c|c|c|c|}
%\hline
 Ref.      &$F_+(0)$  &$F_T(0)$ & $V(0) $   & $A_1(0)$  &  $A_2(0)$  &
$A_0(0)$ &$T_1(0)$ & $T_3(0)$  \\
\hline
This work  &  0.36    &  0.35   & 0.44      &  0.36     &  0.32      & 0.45
& 0.39   &  0.27   \\
SR \cite{colangelo}
           &  0.25    &  -      & 0.47      &  0.37     &  0.40      & 0.30
& 0.38    &  -       \\
Lat+Stech \cite{lat}
           &  -       &  -      & 0.38      &  0.28     &  -         & 0.32
& 0.32    &  -       \\
LCSR'98 \cite{lcsr}
           &  0.34    &  0.374  & 0.46      &  0.34     &  0.28      & 0.47
& 0.38    &  0.26  \\
Lat \cite{ape}
           &  0.30(4) &  0.29(6)&-          &  -        &  -         & -
& -       &  -
\end{tabular}
\end{table}
For the coupling constants we obtain
\begin{eqnarray}
\frac{g_{B_s^*BK}f_V^{(B_s^*)}}{2M_{B_s^*}}= 0.65\pm0.05, \qquad
\frac{g_{B_sBK^*}f_P^{(B_s)}}{2M_{K^*}}=1.65\pm0.1, \qquad
\frac{f_T^{(B_s^*)}}{f_V^{(B_s^*)}}=0.95\pm0.05\qquad
\frac{g_{B_s^*BK^*}}{g_{B_s^*BK}}=1.15\pm0.05. \nonumber
\end{eqnarray}
%\newpage
%b2pi_______________________________________________________________________
%__________________
\subsubsection{$B\to \pi,\rho$}
The $B\to\rho$ transition has been used for determining
the parameters of our quark model in the $u,d$ and $b$ sectors by fitting
the quark-model form factors to
available lattice results on $T_2$ and $A_1$ at large $q^2$ in \cite{mb}.
The corresponding
form factors and the decay rates have been calculated in this article.
We present the results of \cite{mb} in terms of parametrizations for the
form factors of the set (\ref{wsbff}) (see Table \ref{table:fitsb2pi}) which
have not been
given in that paper. The only difference of the results presented here with
the results
of \cite{mb} occurs for the $B\to \pi$ form factors $F_+$ and $F_0$ at
$q^2\ge 20\;GeV^2$. In \cite{mb}
these quantities are not calculated directly but extrapolated from the
region $q^2\le 20\;GeV^2$.
The parametrizations of $f_+$ and $f_-$ in \cite{mb}
correspond to $g_{B^*B\pi}\simeq 50$. The parametrization given here, on the
other hand, corresponds to
$g_{B^*B\pi}\simeq 32$ which is in agreement with recent analyses of this
quantity \cite{mbg,latg}.
At $q^2\le 20\; GeV^2$ both parametrizations describe the results of the
numerical calculation well and
agree with the available results from lattice QCD at $q^2\le 22\;GeV^2$
\cite{ukqcd}.
For these reasons, the earlier result $\Gamma(B\to \pi
l\nu)=8.3|V_{ub}|^2\;ps^{-1}$ calculated with the parametrization of the
form factor $F_+$ from Table \ref{table:fitsb2pi} remains practically
unchanged compared to \cite{mb}.

The form factor $F_0$ at large $q^2$ lies   
below the central lattice values but nevertheless agrees with lattice results 
within the given error bars.
Notice however that in our model the form factor $F_0$ is calculated as a
difference
of $f_+$ and $f_-$ and at large $q^2$ turns out to be much more sensitive to
the subtle details
of the pion wave function, than $f_+$ and $f_-$ separately. A simple
Gaussian wave function
which works quite well for $f_+$ and $f_-$, might not be sufficiently
accurate for $F_0$.
\begin{table}[hbt]
\caption{\label{table:fitsb2pi}The calculated $B \to \pi, \rho$ transition
form factors.
$M_V=M_{B^*}=5.32\; GeV$, $M_P=M_{B}=5.27\; GeV$. 
For the form factors $F_+, F_T, V, A_0, T_1$ the fit formula Eq. (\protect\ref{fit1}) is used, 
for the other form factors - Eq. (\protect\ref{fit3}). }
\centering
\begin{tabular}{|c||c|c|c||c|c|c|c|c|c|c|}
%\hline
 & \multicolumn{3}{c||}{$B \to \pi$} & \multicolumn{7}{c|}{$B \to \rho$}\\
\hline
          &$F_{+}$&$F_{0}$& $F_T$ & $V$   & $A_0$ &  $A_1$  & $A_2$ & $T_1$
& $T_2$  & $T_3$   \\
\hline
$f(0)$     & 0.29 & 0.29  & 0.28  & 0.31  & 0.30  &  0.26  & 0.24  &  0.27
& 0.27   & 0.19    \\
$\sigma_1$ & 0.48 & 0.76  & 0.48  & 0.59  & 0.54  &  0.73  & 1.40  &  0.60
& 0.74   & 1.42    \\
$\sigma_2$ &      & 0.28  &       &       &       &  0.10  & 0.50  &
& 0.19   & 0.51
\end{tabular}
\end{table}
Table \ref{table:compb2pi} compares the results obtained from the quark
model of Ref. \cite{mb} with results from the quark model of Jaus
\cite{jaus}
and latest light-cone sum rule (LCSR) results \cite{lcsr}.
One observes very good agreement between the
quark model of Jaus, LCSRs, and our approach. The only visible
difference with the LCSR method occurs in the form factor $A_0(0)$, which is
caused by small differences of the two methods in $A_1(0)$ and $A_2(0)$
(recall that $A_0(0)=\left((M_1+M_2)A_1(0)-(M_1-M_2)A_2(0)\right)/2M_2$)
This discrepancy  exceeds the 15\% error bar quoted for the LCSR results
only marginally.
If the LCSR results at small $q^2$ and lattice results at large
$q^2$ are correct, our approach surely provides a realistic description of
the
form factors at all kinematically accessible $q^2$ values.
\begin{table}[hbt]
\caption{\label{table:compb2pi}
Comparison of the results of different approaches on weak
$B\to \pi,\rho$ form factors at $q^2=0$.}
\centering
\begin{tabular}{|c||c|c|c|c|c|c|c|c|}
%\hline
 Ref.      &$F_+(0)$  &$F_T(0)$ & $V(0) $   & $A_1(0)$  &  $A_2(0)$  &
$A_0(0)$ &$T_1(0)$ & $T_3(0)$ \\
\hline
This work  &  0.29    &  0.28   & 0.31      &  0.26     &  0.24      & 0.29
& 0.27   &  0.19   \\
Jaus'96 \cite{jaus}
           &  0.27    &  -      & 0.35      &  0.26     &  0.24      & -
&  -     &         \\
LCSR'98 \cite{lcsr}
           &  0.305   &  0.296  & 0.34      &  0.26     &  0.22      & 0.37
& 0.29   &  0.20   \\
Lat \cite{ape}
           &  0.28(4) &  0.28(7)& -         &  -        &  -         & -
& -      &  -
\end{tabular}
\end{table}
Extrapolating the form factors to the poles, we obtain
\begin{eqnarray}
\frac{g_{B^*B\pi}f_V^{(B^*)}}{2M_{B^*}}= 0.6\pm0.05, \qquad
\frac{g_{BB\rho}f_P^{(B)}}{2M_\rho}=1.4\pm0.2, \qquad
\frac{f_T^{(B^*)}}{f_V^{(B^*)}}=0.97\pm0.03, \qquad
\frac{g_{B^*B\rho}}{g_{B^*B\pi}}=1.2\pm0.1. \nonumber
\end{eqnarray}
Using $f_V^{B^*}\simeq 200\;MeV$ gives the estimate
\begin{eqnarray}
g_{BB^*\pi}=32\pm5,\qquad \hat
g=\frac{f_\pi}{2\sqrt{M_BM_{B^*}}}g_{BB^*\pi}= 0.4\pm0.06
\end{eqnarray}

\newpage
The latter value is in good agreement with the lattice result $\hat
g=0.42\pm 0.04\pm0.08$
\cite{latg} and is only slightly smaller than $\hat g=0.5\pm 0.02$ \cite{mbg}
based on combining PCAC with our dispersion approach.

\vspace{.2cm}

Summing up our results on the decays of the nonstrange heavy mesons,
we found no disagreement neither with the exisiting
experimental data nor with the available results of the lattice QCD or
sum rules in their specific regions of validity. The only exception is the
form factor $F_0$
at large $q^2$ in $B\to \pi$ and $D\to \pi$ decays, where our results are lying 
slightly below the lattice points. However, this disagreement can be
related to a strong sensitivity of $F_0$ at large $q^2$ to the details of
the pion wave function.
Small changes in the pion wave function, which only marginally affect $f_+$
and $f_-$, can change the form factor $F_0$. But such subtle effects are beyond the scope
of our present analysis.

In the next section we apply our model to the decays of strange heavy
mesons for which a few new parameters have to be introduced which are
specific to the description of  weak decays of strange heavy mesons
to light mesons.
%\newpage
%___________________________________________________________________________
%________________________
\subsection{Decays of the strange mesons $D_s$ and $B_s$ \label{d}}
Before dealing with these decays, we must first specify the slope parameters
of the
$B_s$ and the $D_s$ wave function. We obtain these parameters by applying
(\ref{fp})  and using
$f_{B_s}/f_B=1.1$ and $f_{D_s}/f_D=1.1$ in agreement with the lattice
estimates for these quantities \cite{lattice}.
The resulting values of the slope parameters are listed in Table
\ref{table:parameters}.
Since all other parameters have already been fixed the
calculation of the form factors is straight forward. The only exceptions are
the decays into the
$\eta, \eta', \phi$ final states. For these decays we need to know the
$\phi$
wave function, the mixing angle and the slope of the radial wave function
of the $s\bar s$ component in $\eta$ and $\eta'$. Our procedure of fixing
these parameters
are discussed in the relevant subsection.

\subsubsection{$D_s\to K, K^*$}
These meson transition are driven by the charged-current $c\to d$ quark
transition.
The results of the calculation are given in Table \ref{table:fitsds2k}.
The predictions
for the semileptonic decay rates are displayed in
Table \ref{table:ratesds2k}.
\begin{table}[hbt]
\caption{\label{table:fitsds2k}The calculated $D_s \to K,K^*$ transition
form
factors. $M_V=M_{D^*}=2.01\; GeV$, $M_P=M_{D}=1.87\; GeV$. 
For the form factors $F_+, F_T, V, A_0, T_1$ the fit formula Eq. (\protect\ref{fit1}) is used, 
for the other form factors - Eq. (\protect\ref{fit3}).}
\centering
\begin{tabular}{|c||c|c|c||c|c|c|c|c|c|c|}
%\hline
 & \multicolumn{3}{c||}{$D_s \to K$} & \multicolumn{7}{c|}{$D_s \to K^*$}\\
\hline
          &$F_{+}$&$F_{0}$& $F_T$ & $V$   & $A_0$ &  $A_1$  & $A_2$ & $T_1$
&   $T_2$  & $T_3$   \\
\hline
$f(0)$     & 0.72 & 0.72  & 0.77  & 1.04  & 0.67  &  0.57  & 0.42  &  0.71
&   0.71   & 0.45    \\
$\sigma_1$ & 0.20 & 0.41  & 0.24  & 0.24  & 0.20  &  0.29  & 0.58  &  0.22
&$-$0.06   & 1.08    \\
$\sigma_2$ &      & 0.70  &       &       &       &  0.42  &       &
&   0.44   & 0.68
\end{tabular}
\end{table}

\begin{table}[hbt]
\caption{\label{table:ratesds2k}
The $D_s\to (K,K^*)l\nu$ decay rates in $10^{10}\;s^{-1}$, $|V_{cd}|=0.22$.}
\centering
\begin{tabular}{|c||c|c|c|c|}
%\hline
Ref & $\Gamma(D_s\to K)$ & $\Gamma(D_s\to K^*)$ & $\Gamma(K^*)/\Gamma(K)$ &
$\Gamma_L/\Gamma_T$ \\
\hline
This work     &  0.63   &  0.38     &   0.6  &  1.21
\end{tabular}
\end{table}
For the coupling constants we obtain
\begin{eqnarray}
\frac{g_{D^*D_sK}f_V^{(D^*)}}{2M_{D^*}}=0.95\pm0.05, \qquad
\frac{g_{DD_sK^*}f_P^{(D)}}{2M_{K^*}}=1.85\pm 0.15,  \qquad
\frac{f_T^{(D^*)}}{f_V^{(D^*)}}=0.9\pm0.1,\qquad
\frac{g_{D^*D_sK^*}}{g_{D^*D_sK}}=1.15\pm0.15.\nonumber
\end{eqnarray}
\newpage
%___________________________________________________________________________
%________________________
\subsubsection{$D_s\to \eta,\eta',\phi$}
These decay modes are induced by the charged current $c\to s$ quark
transition.
The pseudoscalar mesons $\eta$ and $\eta'$ are mixtures of the nonstrange
and the strange components with
the flavour wave functions
$\eta_n\equiv \frac{\bar u u +\bar d d}{\sqrt{2}}$ and $\eta_s=\bar s s$,
respectively,
\begin{eqnarray}
\eta &=&\cos(\varphi)\,\eta_n-\sin(\varphi)\,\eta_s \nonumber\\
\eta'&=&\sin(\varphi)\,\eta_n+\cos(\varphi)\,\eta_s,
\end{eqnarray}
with the angle $\varphi\simeq 40^o$ \cite{amn1,kroll}.
The decay rates of interest are
\begin{eqnarray}
\label{etas}
\Gamma(D_s\to \eta l\nu) &=& \sin^2(\varphi)
\Gamma(D_s\to \eta_s(M_\eta) l\nu)\nonumber\\
\Gamma(D_s\to \eta' l\nu) &=& \cos^2(\varphi)
\Gamma(D_s\to \eta_s(M_{\eta'}) l\nu).
\end{eqnarray}
Let us give a brief explanation of these formulas:
The semileptonic decay rates are determined by the form factor $f_+$.
The spectral representation of this form factor does not involve the
final meson mass explicitly. This means that for the $\bar s s$
component of both
$\eta$ and $\eta'$ we have to deal with the same form factor, which can be
expressed
through the radial wave function of this component.
On the other hand, the phase-space volume of the decay process is determined
by the physical meson masses, as indicated in (\ref{etas}).
It should be clear, however, that the $\eta_s$ is not an
eigenstate of the Hamiltonian and does not have a definite mass.

Assuming universality of the wave functions of the ground state pseudoscalar
$0^-$ nonet,
the radial wave function of the nonstrange component $\Psi_{\eta_n}$
coincides with the
pion radial wave function \cite{amn1}. The radial wave function
$\Psi_{\eta_s}$ should be determined independently.
From the
analysis of a broad set of processes the leptonic decay constant $f_s$ of
the strange component
$\eta_s$,
has been found to lie in the interval $f_s=(1.36\pm 0.04)f_\pi$
\cite{kroll}. This allows us to determine
the slope
parameter $\beta_{\eta_s}$ in such a way that the calculated value of $f_s$
lies in this interval, and the calculated ratio
$\Gamma(D_s \to \eta)/\Gamma(D_s \to \eta')$ agrees with the experimental
data for $\varphi=40^o$.
This procedure yields for the slope parameter $\beta_{\eta_s}=0.45$
\footnote{Another procedure of taking into account the SU(3) breaking
effects to
obtain $\Psi_{\eta_s}$ from $\Psi_{\eta_n}$ has been proposed in
\cite{amn1}.}.
For the slope parameter $\beta_{\phi}$ of the wave function of the
$\phi$-meson,
which is the vector $\bar ss$ state, we expect a value close to
$\beta_{\eta_s}$.

In fact, $\beta_\phi=0.45\;GeV$ leads to the $B_s\to\phi$ transition form
factors which agree well with the LCSR results at $q^2=0$ (see subsection
4).
With all other quark model parameters fixed from the description of the
nonstrange
heavy meson decays and by taking a simple Gaussian form of the radial wave
function,
the decay rate $\Gamma(D_s\to \phi l\nu)$ is a function of $\beta_\phi$.
This function has a minimum at the value $\beta_\phi=0.45\;GeV$;
nevertheless, the corresponding value of the decay rate is $1\sigma$ above
the central
experimental value).

The results of our calculations are given in Tables \ref{table:fitsds2phi}
and \ref{table:ratesds2ss}.
\begin{table}[hbt]
\caption{\label{table:fitsds2phi}The calculated $D_s \to \eta_s,\phi$
transition form
factors. $M_V=M_{D_s^*}=2.11\; GeV$, $M_P=M_{D_s}=1.97\; GeV$. 
For the form factors $F_+, F_T, V, A_0, T_1$ the fit formula Eq. (\protect\ref{fit1}) is used, 
for the other form factors - Eq. (\protect\ref{fit3}).}
\centering
\begin{tabular}{|c||c|c|c|c|c||c|c|c|c|c|c|c|}
%\hline
 &   \multicolumn{3}{r}{$D_s \to \eta_s(M_\eta)$}
 &   \multicolumn{2}{r||}{$D_s \to \eta_s(M_{\eta'})$}
 &   \multicolumn{7}{c|}{$D_s \to\phi$}\\
\hline
            &$F_{+}$&$F_{0}$&$F_{T}$&$F_{0}$& $F_T$
     & $V$   & $A_0$ &  $A_1$  & $A_2$ & $T_1$ &   $T_2$  & $T_3$   \\
\hline
$f(0)$      & 0.78  & 0.78  & 0.80  & 0.78  & 0.94
            &  1.10 &  0.73 &  0.64 & 0.47  &  0.77   &   0.77   & 0.46
\\
$\sigma_1$ & 0.23  & 0.33  & 0.24  & 0.21  & 0.24
            &  0.26 &  0.10 &  0.29 &  0.63   & 0.25  &   0.02   & 1.34
\\
$\sigma_2$ &       & 0.38  &       & 0.76  &
            &       &       &       &       &         &   2.01   & 0.45
\end{tabular}
\end{table}
\begin{table}[hbt]
\caption{\label{table:ratesds2ss}
The $D_s\to (\eta,\eta',\phi)l\nu$ decay rates in $10^{10}\;s^{-1}$,
$|V_{cs}|=0.975$.
The experimental rates are obtained from the corresponding branching ratios
using the
$D_s$ lifetime $\tau_{D_s}=0.495\pm 0.013\;ps$ from the 1999 update
\protect\cite{pdg}}
\centering
\begin{tabular}{|c||c|c|c|}
%\hline
Ref & $\Gamma(D_s\to \eta)$ & $\Gamma(D_s\to \eta')$ & $\Gamma(D_s\to \phi)$
\\
\hline
This work     &  5.0            &  1.85           &   5.1    \\
Exp \cite{pdg}&  5.2$\pm$1.3   &  2.0$\pm$0.8   &   4.04$\pm$1.01
\end{tabular}
\end{table}
For the coupling constants we obtain
\begin{eqnarray}
\frac{g_{D_s^*D_s\eta_s}f_V^{(D_s^*)}}{2M_{D_s^*}}=1.0\pm0.1, \qquad
\frac{g_{D_sD_s\phi}f_P^{(D_s)}}{2M_{\phi}}=1.6\pm 0.3,  \qquad
\frac{f_T^{(D_s^*)}}{f_V^{(D_s^*)}}=0.93\pm0.03,\qquad
\frac{g_{D_s^*D_s\phi}}{g_{D_s^*D_s\eta_s}}=1.08\pm0.04.
\nonumber
\end{eqnarray}

%newpage
%________________________________________________________________________
\subsubsection{$B_s\to K,K^*$}
This mode is driven by the $b\to u$ charged current transition.
The only additional new parameter needed here is the slope of the $B_s$ wave
function. We
obtain it by using (\ref{fp}) and taking $f_{B_s}/f_B=1.1$. The results of
our
calculation are given in Table \ref{table:fitsbs2k}.
\begin{table}[hbt]
\caption{\label{table:fitsbs2k}
The calculated $B_s \to K, K^*$ transition form factors. $M_V=M_{B^*}=5.32\;
GeV$, $M_P=M_{B}=5.27\; GeV$. 
For the form factors $F_+, F_T, V, A_0, T_1$ the fit formula Eq. (\protect\ref{fit1}) is used, 
for the other form factors - Eq. (\protect\ref{fit3}).}
\centering
\begin{tabular}{|c||c|c|c||c|c|c|c|c|c|c|}
%\hline
 & \multicolumn{3}{c||}{$B_s \to K$} & \multicolumn{7}{c|}{$B_s \to K^*$}\\
\hline
          &$F_{+}$&$F_{0}$& $F_T$ & $V$   & $A_0$ &  $A_1$  & $A_2$ & $T_1$
&   $T_2$  & $T_3$   \\
\hline
$f(0)$     & 0.31 & 0.31  & 0.31  & 0.38  & 0.37  &  0.29  & 0.26  &  0.32
&   0.32   & 0.23    \\
$\sigma_1$ & 0.63 & 0.93  & 0.61  & 0.66  & 0.60  &  0.86  & 1.32  &  0.66
&   0.98   & 1.42    \\
$\sigma_2$ & 0.33 & 0.70  & 0.30  & 0.30  & 0.16  &  0.60  & 0.54  &  0.31
&   0.90   & 0.62
\end{tabular}
\end{table}
These form factors lead to the following relations
\begin{eqnarray}
\frac{g_{B^*B_sK}f_V^{(B^*)}}{2M_{B^*}}=0.44\pm0.04, \qquad
\frac{g_{BB_sK^*}f_P^{(B)}}{2M_{K^*}}=1.3\pm 0.1, \qquad
\frac{f_T^{(B^*)}}{f_V^{(B^*)}}=0.95\pm0.05,\qquad
\frac{g_{B^*B_sK^*}}{g_{B^*B_sK}}=1.2\pm0.1.
\nonumber
\end{eqnarray}
The form factors at $q^2=0$ are compared with the LCSR predictions in Table
\ref{table:compbs2k}.
We observe some disagreement between our predictions and the LCSR
calculation which
gives smaller values for all the form factors. A closer look at the origin
of this discrepancy shows
that its source is the strength and sign of the SU(3)-breaking effects.
They lead to opposite
corrections in the two approaches.
\begin{table}[h]
\caption{\label{table:compbs2k}
Comparison of the QM and LCSR results on the $B_s\to K,K^*$ form factors at
$q^2=0$.}
\centering
\begin{tabular}{|c||c|c|c|c|c|c|}
%\hline
 Ref.        & $V(0) $   & $A_1(0)$  &  $A_2(0)$  & $A_0(0)$ &$T_1(0)$ &
$T_3(0)$ \\
\hline
This work    & 0.38      &  0.29     &  0.26      & 0.37     & 0.32    &
0.23   \\
LCSR'98 \cite{lcsr}
             & 0.262     &  0.19     &  0.164     & 0.254    & 0.22    &
0.16
\end{tabular}
\end{table}
To discuss these  SU(3) breaking effects, let us start with
$B\to \rho$, which in fact differs from the $B_s\to K^*$ only
by the flavour of the spectator quark, and move to $B_s\to K^*$ by
accounting
for the SU(3) violating effects:

Within the LCSR method there are two changes which affect the form factors:
first, the change $f_{B_s}\to f_B$
leads to an increase of the $B_s\to K^*$ form factors;
second, the change of the symmetric twist-two distribution amplitude of the
$\rho$-meson
to the asymmetric one of the $K^*$ meson
leads to a decrease of the form factors. The second
effect turns out to be much stronger than the first one with the result of
an overall decrease of the
form factors.

In the quark model, the same SU(3) breaking effects take place:
The change of the spectator mass (it determines the increase of
$f_{B_s}/f_{B}$)
and the change of the $K^*$ meson wave function (due to the change of both
the quark mass
and the slope parameter of the light meson wave function). Here the
influence of the slope of the heavy
meson wave function upon the form factor is only marginal. Therefore, the
resulting effect of these
changes leads to an increase of the form factors.

We want to point out that the difference between the results of the two
approaches
does not arise from specific effects (higher twists, higher radiative
corrections etc) which are present in the LCSRs but absent in the quark
model.
The observed difference is only due to the different strength of the SU(3)
violating
effects at the level of the twist-2 distribution amplitude. As was discussed
in \cite{amn},
this distribution amplitude can be expressed through the radial soft wave
function of the meson.
The change of the quark-model wave function caused by SU(3) violating
effects does not induce a strong
asymmetry in the leading twist-2 distribution amplitude.

In view of the discrepancy between our results and the LCSR it would be
interesting
to have independent calculations of the $B_s\to K^*$ form factors at small
$q^2$ from the
3-point sum rules, as well as a lattice calculation for large $q^2$.
%___________________________________________________________________________
%________________
\subsubsection{$B_s\to \eta,\eta',\phi$}
These weak meson transitions are induced by the FCNC $b\to s$ quark
transition.
The results of the form factor calculation are given in Table
\ref{table:fitsbs2phi} and compared with the
LCSR predictions at $q^2=0$ in Table \ref{table:compbs2phi}. The agreement
between the two values is satisfactory at least within the declared 15\%
accuracy of the LCSR predictions.
This allows us to expect that also the $D_s\to \phi, \eta,\eta'$ form
factors and the corresponding
decay rates (given earlier in subsection 2) are calculated reliably.
%\newpage
\begin{table}[hbt]
\caption{\label{table:fitsbs2phi}The calculated $B_s \to \eta_s,\phi$
transition form
factors. $M_V=M_{B_s^*}=5.42\; GeV$, $M_P=M_{B_s}=5.37\; GeV$. 
For the form factors $F_+, F_T, V, A_0, T_1$ the fit formula Eq. (\protect\ref{fit1}) is used, 
for the other form factors - Eq. (\protect\ref{fit3}).}
\centering
\begin{tabular}{|c||c|c|c|c|c||c|c|c|c|c|c|c|}
%\hline
 &   \multicolumn{3}{r}  {$B_s \to \eta_s(M_\eta)$}
 &   \multicolumn{2}{r||}{$B_s \to \eta_s(M_{\eta'})$}
 &   \multicolumn{7}{c|} {$B_s \to\phi$}\\
\hline
            &$F_{+}$&$F_{0}$&$F_{T}$&$F_{0}$& $F_T$
     & $V$   & $A_0$ &  $A_1$  & $A_2$ & $T_1$ &   $T_2$  & $T_3$   \\
\hline
$f(0)$      & 0.36  & 0.36  & 0.36  & 0.36  & 0.39
            &  0.44 &  0.42 &  0.34 & 0.31  &  0.38   &   0.38   & 0.26
\\
$\sigma_1$  & 0.60  & 0.80  & 0.58  & 0.80  & 0.58
            &  0.62 &  0.55 &  0.73 &  1.30   & 0.62  &   0.83   & 1.41
\\
$\sigma_2$  & 0.20  & 0.40  & 0.18  & 0.45  & 0.18
            &  0.20 &  0.12 &  0.42 &  0.52 &  0.20   &   0.71   & 0.57
\end{tabular}
\end{table}
\begin{table}[h]
\caption{\label{table:compbs2phi}
Comparison of the QM and LCSR results on the $B_s\to \phi$ form factors at
$q^2=0$.}
\centering
\begin{tabular}{|c||c|c|c|c|c|c|}
%\hline
 Ref.        & $V(0) $   & $A_1(0)$  &  $A_2(0)$  & $A_0(0)$ &$T_1(0)$ & $T_3(0)$ \\
\hline
This work    & 0.44      &  0.34     &  0.31      & 0.42     & 0.38    &  0.26   \\
LCSR'98 \cite{lcsr}
             & 0.433     &  0.296    &  0.255     & 0.382    & 0.35    &  0.25
\end{tabular}
\end{table}
For the coupling constants we obtain
\begin{eqnarray}
\frac{g_{B_s^*B_s\eta_s}f_V^{(B_s^*)}}{2M_{B_s^*}}=0.6\pm0.05, \qquad
\frac{g_{B_sB_s\phi}f_P^{(B_s)}}{2M_{\phi}}=1.5\pm 0.1, \qquad
\frac{f_T^{(B_s^*)}}{f_V^{(B_s^*)}}=0.95\pm0.05, \qquad
\frac{g_{B_s^*B_s\phi}}{g_{B_s^*B_s\eta_s}}=1.13\pm0.06.\nonumber
\end{eqnarray}
\section{conclusion}
We have calculated numerous form factors of heavy meson transitions to light
mesons which
are relevant for the semileptonic (charged current) and penguin
(flavor-changing neutral current)
decay processes. Our approach is based on evaluating the triangular decay
graph within a relativistic
quark model which has the correct analytical properties and satisfies
all known general requirements of QCD.

The model connects different decay channels in a unique way and gives the
form factors for all
relevant $q^2$ values. The disadvantage of the constituent quark model
connected with its dependence on
ill-defined parameters such as the effective constituent quark masses, have
been reduced by
using several constraints: the quark masses and the slope parameters of the
wave functions are
chosen such that the calculated form factors reproduce the lattice results
for the $B\to \rho$
form factors at large $q^2$ and the observed integrated rates of the
semileptonic $D\to K,K^*$ decays.

Our main results are as follows:
\begin{itemize}
\item
In spite of the rather different masses and properties of mesons
involved in weak transitions, all existing data on the form factors, both
from
theory and experiment, can be understood in our quark picture.
Namely, all the form factors are essentially described by the few degrees of
freedom of constituent quarks, i.e. their wave functions and their effective
masses. Details of the soft wave functions are not crucial; only the
spatial extention  of these wave functions of order of the confinement scale
is important.
In other words, only the meson radii are essential.

\item
The calculated transition form factors are in all cases in good agreement
with the results available from lattice QCD and from sum rules
in their specific regions of validity. The only exception is a disagreement
with the LCSR results for the $B_s\to K^*$ transition where
we predict larger form factors. This disagreement is caused by a different
way of taking into account the
SU(3) violating effects when going from $B\to\rho$ to $B_s\to K^*$ and is
not related to specific
details of the dynamics of  the decay process.
We suspect that the LCSR method overestimates the SU(3) breaking
in the long-distance region
but this problem deserves further clarification. 

\newpage
\item
We have estimated the products  
of the meson weak and strong coupling constants by using the fit formulas for 
the form factors for the extrapolation to the meson pole. 
The error of such estimates connected with the errors in the extrapolation 
procedure is found to be around 5-10\%. 
\end{itemize}
We cannot provide for definite error estimates of our predictions for the form factors 
because of the approximate
character of the constituent quark model. However from the fine agreement
obtained in cases where checks are possible,
we believe that the actual accuracy of our predictions for the form factors
is around 10\%.
Since some parameters have been fixed by using lattice results and have also
been tested
using the sum rule predictions, further improvements of the accuracy of our
predictions will follow
if these approaches attain smaller errors.
Of course, each precisely measured decay will also allow a more accurate
determination of the
parameters of the model and thus can be used to diminish the errors at least
for
closely related decays.

%\newpage
\section{Acknowledgments}
We dedicate this paper to Franz Wegner on the occasion of his 60-th birthday. 
We are grateful to D. Becirevic, M. Beyer, Th. Feldmann, A. Le Yaouanc, N.
Nikitin, O. P\`ene, and S. Simula
for helpful and interesting discussions. One of us (DM) gratefully
acknowledges
the financial support of the Alexander von Humboldt Stiftung.

\section{Appendix: Weak and strong meson coupling constants}
We provide here definitions of the coupling constants which determine the
behavior
of the form factors at $q^2$ near the resonance poles (beyond the decay
region).
We consider as an example the case of the $B\to \pi,\rho$ transition.

\vspace{.6cm}
\centerline{\small \it 1. Weak decay constants}
\vspace{.4cm}

Weak decay constants of mesons are determined by the following relations
\begin{eqnarray}
\langle B(q)|\bar b(0)\gamma_\mu\gamma_5 q(0) |0 \rangle &=&i
f_P^{(B)}q_\mu\nonumber\\
\langle B^*(q)|\bar b(0)\gamma_\mu q(0) |0 \rangle
&=&\epsilon_\mu^{(B^*)}M_{B^*}f_V^{(B^*)}
\nonumber\\
\langle B^*(q)|\bar b(0)\sigma_{\mu\nu} q(0) |0 \rangle &=&
i(\epsilon_\mu^{(B^*)}q_\nu-\epsilon_\nu^{(B^*)}q_\mu)f_T^{(B^*)},
\end{eqnarray}
where $\epsilon_\mu^{(B^*)}$ is the $B^*$ polarization vector.
In the heavy quark limit one has $f_P^{(B)}=f_V^{(B^*)}=f_T^{(B^*)}$.

\vspace{.6cm}
\centerline{\small \it 2. Strong coupling constants}
\vspace{.4cm}

Strong coupling constants are connected with the three-meson amplitudes as
follows
\begin{eqnarray}
\langle \pi(p_2)B(p_1)|B^*(q) \rangle
&=&\frac{1}{2}g_{B^*B\pi}P_\mu\epsilon_\mu^{(B^*)}
\nonumber\\
\langle \rho(p_2)B(p_1)|B^*(q) \rangle
&=&\frac{1}{2}\epsilon_{\alpha\beta\mu\nu}
\epsilon_\alpha^{(\rho)}\epsilon_\beta^{(B^*)}P_\mu q_\nu
\frac{g_{B^*B\rho}}{M_B^*}
\nonumber\\
\langle \rho(p_2)B(p_1)|B^*(q) \rangle
&=&\frac{1}{2}g_{B^*B\rho}P_\mu\epsilon_\mu^{(\rho)},
\end{eqnarray}
where $q=p_1-p_2$, $P=p_1+p_2$ and $\epsilon_\mu$ is the polarization vector
of the
vector meson.

In the heavy quark limit $g_{B^*B\rho}=g_{BB\rho}$.

\vspace{.6cm}
\centerline{\small \it 3. Form factors}
\vspace{.4cm}

The form factors $F_+$, $F_T$, $V$, $T_1$
contain pole at $q^2=M^2_{B^*}$ due to the contribution of the intermediate
$B^*$ ($1^-$ state) in the $q^2$ channel. The residue of this pole is given
in terms of the
product of the weak and strong coupling constants such that in the region
$q^2\simeq M_{B^*}^2$ the form factors read
\begin{eqnarray}
F_+&=&\frac{g_{B^*B\pi}f_V^{(B^*)}}{2M_B^*}\frac{1}{1-q^2/M^2_{B^*}}+...,
\nonumber\\
F_T&=&\frac{g_{B^*B\pi}f_T^{(B^*)}}{2M_B^*}\frac{M_B+M_\pi}{M_B^*}\frac{1}{1
-q^2/M^2_{B^*}}+...,\nonumber\\
V&=&\frac{g_{B^*B\rho}f_V^{(B^*)}}{2M_B^*}\frac{M_B+M_\rho}{M_{B^*}}\frac{1}
{1-q^2/M^2_{B^*}}+...,\nonumber\\
T_1&=&\frac{g_{B^*B\rho}f_T^{(B^*)}}{2M_B^*}\frac{1}{1-q^2/M^2_{B^*}}+...
\nonumber\\
\end{eqnarray}
Here $\dots$ stand for the terms non-singular at $q^2=M_{B^*}^2$.

Similarly, $A_0$ contains the contribution of the $B$ ($0^-$ state).
In the region of
$q^2\simeq M_B^2$  it can be represented as follows
\begin{eqnarray}
A_0=\frac{g_{BB\rho}f_{P}^{(B)}}{2M_B}\frac{M_B}{2M_\rho}\frac{1}{1-q^2/M^2_
{B}}+...
\end{eqnarray}
First let us notice that the residues of the form factors are not all
independent and are
connected with each other as follows:
\begin{eqnarray}
\label{constraint}
\frac{Res(F_T)Res(V)}{Res(F_+)Res(T_1)}=\frac{M_B+M_\rho}{M_{B^*}}\frac{M_B+
M_\pi}{M_{B^*}}.
\end{eqnarray}
This relation can be used as a cross-check of the consistency of the
extrapolation for the form factors.

The coupling constants are related to the residues of
the form factors according to the relations
\begin{eqnarray}
\frac{g_{B^*B\pi}f_V^{(B^*)}}{2M_{B^*}}=Res(F_+)&&,\qquad \qquad
\frac{g_{BB\rho}f_P^{(B)}}{2M_{\rho}}=2Res(A_0),\nonumber\\
\frac{f_T^{(B^*)}}{f_V^{(B^*)}}=\frac{Res(F_T)}{Res(F_+)}\frac{M_{B^*}}{M_B+
M_\pi}&&,\qquad \qquad
\frac{g_{B^*B\rho}}{g_{B^*B\pi}}=\frac{Res(V)}{Res(F_+)}\frac{M_{B^*}}{M_B+M
_\rho}.
\end{eqnarray}


\begin{thebibliography}{99}
\bibitem{expd2pi}The rates for the $D\to \pi$ and $D\to\rho$ are obtained by
combining
the values for the $D\to K$ and $D\to K^*$ with the measurements of the
ratios
of the branching fractions $D\to \pi/D\to K$ by CLEO \cite{cleopik} and
$D\to \rho/D\to K^*$ by E653 \cite{e653}.
\end{thebibliography}
\end{document}